\documentclass[twocolumn,twocolappendix]{aastex631}

\begin{document}

\title{Comprehensive analysis of a symbiotic candidate V503~Her}

\author[0000-0001-6355-2468]{Jaroslav Merc}
\affiliation{Astronomical Institute, Faculty of Mathematics and Physics, Charles University, 
V Hole\v{s}ovi\v{c}k{\'a}ch 2, 180 000, Prague, Czech Republic}
\correspondingauthor{Jaroslav Merc}
\email{jaroslav.merc@mff.cuni.cz}

\author[0000-0003-4299-6419]{Rudolf G{\'a}lis}
\affiliation{Institute of Physics, Faculty of Science, P. J. \v{S}af{\'a}rik University, 
Park Angelinum 9, 040 01 Ko\v{s}ice, Slovak Republic}

\author[0000-0002-4387-6358]{Marek Wolf}
\affiliation{Astronomical Institute, Faculty of Mathematics and Physics, Charles University, 
V Hole\v{s}ovi\v{c}k{\'a}ch 2, 180 000, Prague, Czech Republic}

\author[0000-0001-5541-2836]{Pavol A. Dubovský}
\affiliation{Vihorlat Astronomical Observatory, Mierová 4, 066 01 Humenné, Slovak Republic}

\author[0000-0002-1012-7203]{Jan K\'{a}ra}
\affiliation{Astronomical Institute, Faculty of Mathematics and Physics, Charles University, 
V Hole\v{s}ovi\v{c}k{\'a}ch 2, 180 000, Prague, Czech Republic}

\author{Forrest Sims}
\affiliation{Astronomical Ring for Amateur Spectroscopy Group (ARAS)}

\author{James R. Foster}
\affiliation{Astronomical Ring for Amateur Spectroscopy Group (ARAS)}

\author{Tom\'{a}\v{s} Medulka}
\affiliation{Vihorlat Astronomical Observatory, Mierová 4, 066 01 Humenné, Slovak Republic}

\author{Christophe Boussin}
\affiliation{Astronomical Ring for Amateur Spectroscopy Group (ARAS)}

\author{John P. Coffin}
\affiliation{Astronomical Ring for Amateur Spectroscopy Group (ARAS)}

\author{Christian Buil}
\affiliation{Astronomical Ring for Amateur Spectroscopy Group (ARAS)}

\author{David Boyd}
\affiliation{Astronomical Ring for Amateur Spectroscopy Group (ARAS)}

\author{Jacques Montier}
\affiliation{Astronomical Ring for Amateur Spectroscopy Group (ARAS)}



\begin{abstract}
V503~Her was previously proposed as an eclipsing symbiotic candidate based on photometric behavior and spectroscopic appearance indicating the composite optical spectrum. To investigate its nature, we analyzed long-term photometric observations covering one hundred years of its photometric history and new low-resolution optical spectroscopic data, supplemented with the multifrequency measurements collected from several surveys and satellites. Based on the analysis presented in this paper, we claim that V503~Her is not an eclipsing binary star. The optical and infrared wavelengths are dominated by a K-type bright giant with an effective temperature of 4\,500\,K, luminosity of 1\,900\,L$_\odot$, and sub-solar metallicity on the asymptotic giant branch showing semiregular complex multi-periodic pulsation behavior. V503~Her does not show the characteristics of strongly interacting symbiotic variables, but some pieces of evidence suggest that it could still be one of the 'hidden' accreting-only symbiotic systems. However, the currently available data do not allow us to fully confirm or constrain the parameters of a possible companion.

\end{abstract}

\keywords{Symbiotic binary stars (1674) --- Evolved stars (481) --- Eclipsing binary stars (444) --- Asymptotic giant branch stars (2100) --- Pulsating variable stars (1307)\vspace{5mm}}


\section{Introduction} \label{sec:intro}

Symbiotic stars are interacting binary systems that typically consist of a cool giant of spectral type M (less frequently K), a hot white dwarf (or a neutron star), and a~complex circumbinary nebula. The cool giants dominate the IR part of the spectrum, the hot components prevail in the UV region and further towards shorter wavelengths. The nebular radiation is most prominent in the optical, and it also comprises strong emission lines \citep[see, e.g., the reviews by][]{2012BaltA..21....5M, 2019arXiv190901389M}. 

While symbiotics are unique laboratories for studying several astrophysical processes and stellar evolution, the number of known systems in the Milky Way is still much lower than any estimate of the symbiotic population size. In recent years, several surveys have focused on discovering new symbiotic stars, both in the Milky Way \citep[e.g.,][]{2008A&A...480..409C,2010A&A...509A..41C,2009MNRAS.395.1121K,2013MNRAS.432.3186M, 2014MNRAS.440.1410M,2014A&A...567A..49R,2021MNRAS.505.6121M,2021MNRAS.502.2513A,2023MNRAS.519.6044A} and in the Local Group \citep[e.g.,][]{2008MNRAS.391L..84G,2015MNRAS.447..993G, 2014MNRAS.444..586M, 2017MNRAS.465.1699M,2018arXiv181106696I,2019A&A...627A.128S}. However, many of the candidates accumulated over the years have only been very poorly studied, and several are included in this category of objects, in fact, incorrectly \citep[see, e.g., the case of LAMOST J202629.80+423652.0 in][]{2020CoSka..50..672A}. Based on our New Online Database of Symbiotic Variables \citep[][]{2019RNAAS...3...28M}, we have initiated an observational campaign to analyze some of the supposed but poorly characterized symbiotic stars and candidates \citep[see also][]{2020MNRAS.499.2116M,2021MNRAS.506.4151M,2022MNRAS.510.1404M,2023MNRAS.523..163M}.

One of the targets we selected for the analysis was V503~Her. The object was first mentioned as a possible symbiotic star by \citet{1980VeSon...9..197M} but has never been studied in detail since then. It was also included, as a~suspected symbiotic binary, in the book by \citet[][]{1986syst.book.....K}, as well as in the catalogs by \citet{2000A&AS..146..407B} and \citet{2019ApJS..240...21A}. At the time of writing of this article, V503~Her was denoted as a confirmed symbiotic star in the SIMBAD database, with reference to the General Catalogue of Variable Stars \citep[GCVS,][]{2017ARep...61...80S}. GCVS itself listed the star only as a possible symbiotic binary (type 'ZAND:'). 

There is only limited observational data for V503~Her available in the literature. \citet{1978PASP...90..526B} obtained a spectrum of the object using the spectrograph mounted on the Kitt Peak 2.1-m telescope while searching for bright quasi-stellar objects. According to the author, the spectrum showed 'TiO bands about as well-developed as in type M2, but very weak'. In addition, he noted that the blue part of the spectrum 'appears filled in, as if by a~hotter star' and suggested that the spectrum might be composite. Up to our knowledge, the only subsequent spectrum of V503~Her was published by \citet{2002A&A...383..188M}. Their spectrum showed a continuum similar to that of a K-type star with a hint of TiO bands and no emission lines.

In addition to the spectroscopic observations, \citet{1980VeSon...9..197M} analyzed the photometric variability of V503~Her based on Sonnenberg plates, suggesting long-term variability similar to symbiotic stars and also fluctuations with varying amplitudes on timescales of 80-100 days. More recent photometric observations obtained by the All Sky Automated Survey (ASAS) were analyzed by \citet{2013AcA....63..405G}, who suggested that V503~Her might be an eclipsing symbiotic star with an orbital period of around 1\,575 days, based on two detected minima at JD 2\,453\,145 and JD 2\,454\,720. They also recorded pulsations with a~period of 130 $\pm$ 2 days.

In this paper, we present the results of the analysis of our new spectroscopic observations and photometry collected from various surveys, supplemented by the available information on V503~Her from the literature. On the basis of our results, we claim that the object is not an eclipsing system, while it might be an accreting-only symbiotic binary. The paper is organized as follows: in Section \ref{sec:observations}, we describe the used observational data, and in Section \ref{sec:results}, we discuss the results regarding the photometric variability, spectroscopic appearance, stellar parameters, and classification of V503~Her as a symbiotic star. Finally, Section \ref{sec:conclusions} concludes and summarizes our findings. 

\section{Observational data}\label{sec:observations}

The spectroscopic observations of V503~Her were obtained in cooperation with the ARAS Group\footnote{https://aras-database.github.io/database/symbiotics.html} \citep[\textit{Astronomical Ring for Amateur Spectroscopy};][]{2019CoSka..49..217T}, which is an initiative dedicated to the promotion of amateur astronomical spectroscopy and pro/am collaboration. Between JD\,2\,456\,448 and JD\,2\,459\,464 (June 5, 2013 - April 2, 2023), we have collected 23 low-resolution optical spectra of V503~Her ($R \sim 600 - 1\,100$). Most of them were obtained in the scope of our observational campaign initiated in February 2021. The log of observations is available in Tab. \ref{table:log_obs} in the Appendix \ref{app:log}.

For the analysis of long-term photometric variations of V503~Her, we collected the available photometry obtained from the Northern Sky Variability Survey \citep[NSVS; ][]{2004AJ....127.2436W}, the All-Sky Automated Survey \citep[ASAS;][]{1997AcA....47..467P}, the Super Wide Angle Search for Planets \citep[SuperWASP;][]{2010A&A...520L..10B}, the All-Sky Automated Survey for Supernovae \citep[ASAS-SN;][]{2014ApJ...788...48S, 2017PASP..129j4502K}, and the Zwicky Transient Facility survey \citep[ZTF;][]{2019PASP..131a8003M}. These were supplemented by the observations obtained from the database of the American Association of Variable Star Observers\footnote{https://www.aavso.org/} \citep[AAVSO; ][]{kafka}. We also used observations from the DASCH (Digital Access to a Sky Century at Harvard) archive of digitized photographic plates from the Harvard College Observatory \citep{2010AJ....140.1062L}. Only unflagged data points were used for analysis. The observations obtained from Sonnenberg plates, collected and analyzed by \citet[][]{1980VeSon...9..197M} were also included in the light curve. The list of sources of photometric data, together with filters and dates, is available in the Tab. \ref{table:log_obs_photometric} in the Appendix \ref{app:log}.

To search for short-term photometric variability of V503 Her, we executed a 120-min observing run at JD\,2\,460\,040.5 (April 6, 2023) in Johnson \textit{B} filter using the Mayer 65-cm telescope in Ondřejov, Czech Republic. In total, 38 frames were obtained with an exposure time of 180\,s, each.

To construct the multi-frequency spectral energy distribution (SED) of V503~Her, we have collected data from the Galaxy Evolution Explorer satellite \citep[GALEX;][]{2017ApJS..230...24B}, the \textit{Gaia} EDR3/DR3 \citep[][]{2021A&A...649A...1G,2022arXiv220800211G}, the Panoramic Survey Telescope and Rapid Response System \citep[Pan-STARRS;][]{2020ApJS..251....7F}, the AAVSO Photometric All-Sky Survey \citep[APASS;][]{2015AAS...22533616H}, the Two Micron All-Sky Survey \citep[2MASS;][]{2006AJ....131.1163S}, and the  \textit{Wide-field Infrared Survey Explorer} \citep[\textit{WISE};][]{2010AJ....140.1868W}. The used filters and observing dates are summarized in the Tab. \ref{table:log_obs_sed} in the Appendix \ref{app:log}.

\section{Analysis, results and discussion}\label{sec:results}
In this section, we describe the process of analysis, and discuss the results concerning the photometric and spectroscopic behavior of V503~Her, its parameters and evolutionary state, and the previously proposed symbiotic classification of this object.

\subsection{Distance and reddening}\label{sec:distance}

In order to investigate not only the proposed symbiotic nature of V503~Her but also some of its parameters, it is necessary to adopt a value of its distance. Although the parallaxes obtained by \textit{Gaia} have some limitations \citep{2021A&A...649A...2L,2021ApJ...907L..33S}, we have chosen to adopt the distance estimate provided by \citet{2021AJ....161..147B}. They employed a probabilistic approach utilizing the \textit{Gaia} EDR3 measurements (parallax, color, and apparent magnitude of the star) and a 3D model of the Galaxy. For V503~Her, whose parallax is $(0.077 \pm 0.014)$\,mas in the \textit{Gaia} EDR3 \citep[][]{2021A&A...649A...1G}, they obtained a distance of 10.6\,kpc (with an uncertainty of 9.2 - 12.5\,kpc). We use this distance estimate for our subsequent analyses.

The total Galactic extinction in the direction of V503~Her, as given by the map of \citet{2011ApJ...737..103S}, is $E(B-V) = 0.07$\,mag. Using the 3D extinction map of \citet[][]{2019ApJ...887...93G} and the adopted distance of V503~Her, a similar value of $E(B-V) = 0.09$\,mag can be obtained. Therefore, we have chosen to adopt an extinction value of $E(B-V) = 0.08$\,mag to deredden the observed fluxes and spectra presented in this study. We note that the reddening estimates assume a total-to-selective absorption ratio of $R_{\rm V} = 3.1$ and follow the reddening law of \citet{1989ApJ...345..245C}.

\subsection{Photometric variability} \label{sec:long-term}

\begin{figure*}
	\plotone{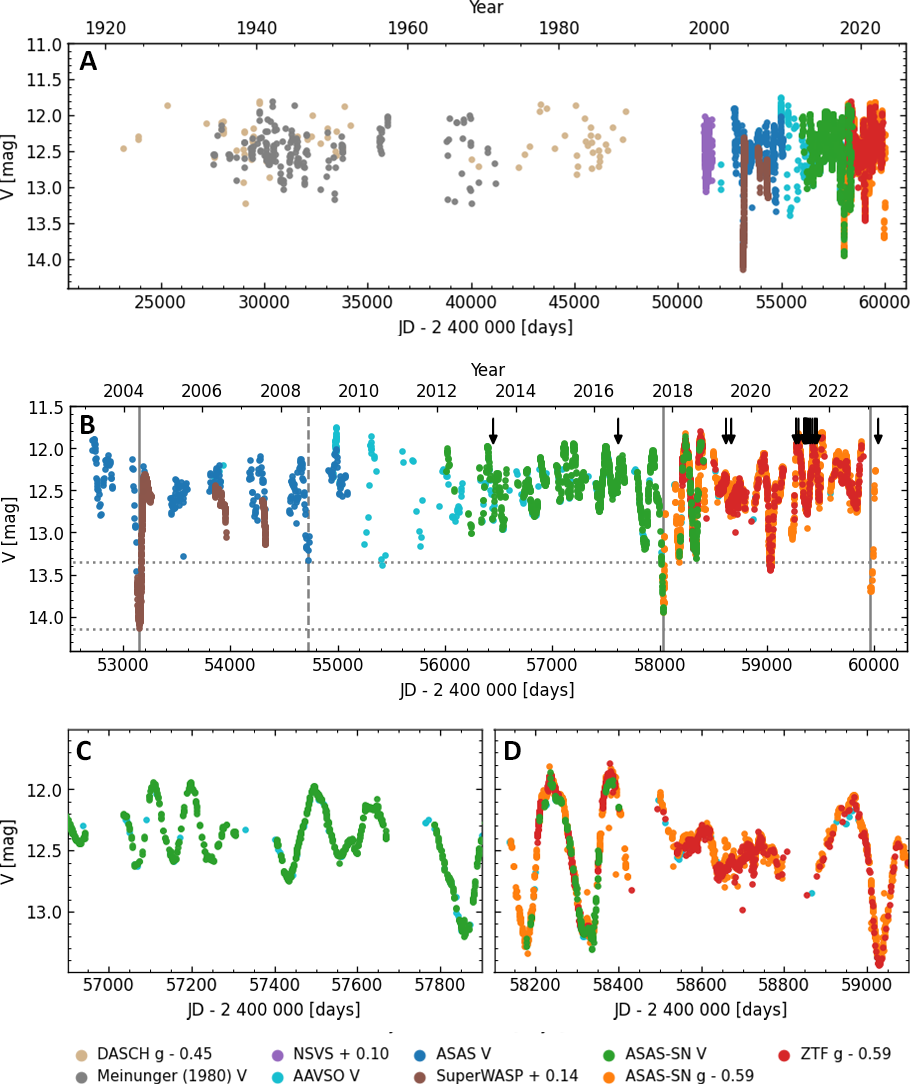}
    \caption{Photometric data on V503 Her. \textbf{A:} Historical light curve of V503~Her, covering the period of the years 1922 - 2023, constructed on the basis of data from DASCH, Sonnenberg plates \citep[][]{1980VeSon...9..197M}, NSVS, AAVSO, ASAS, SuperWASP, ASAS-SN, and ZTF. \textbf{B:} Part of the V503~Her light curve (2002 - 2023). The solid vertical  gray lines mark the position of deep minima discussed in the text, while the dashed vertical line denotes the minimum considered an eclipse by \citet{2013AcA....63..405G}. The horizontal dotted lines show the level of the brightness of the object in the middle of the minima in 2004 and 2009, respectively. Arrows denote the times when our spectra were obtained. \textbf{C\&D:} Changes in the pulsation amplitude and timescale in the light curve of V503~Her.}
    \label{fig:lc_full}
\end{figure*}

To study the long-term behavior of V503~Her, we compiled data collected in multiple bands that span the years 1922 to 2023 from various sources, including photographic plates, all-sky surveys, and observations of amateur observers. We linearly shifted all individual datasets to the $V$ magnitude level to construct a single composite light curve. We determined the offset values for each dataset by analyzing a sample of nine field stars showing no significant variability in the ASAS-SN and ZTF surveys, distributed in the $(B-V)$ color range of -0.075 to 1.325 mag \citep{2015AAS...22533616H} around the color index of V503~Her. We adopted \mbox{$(B-V) = 1.309$\,mag} for V503~Her, which is the median value based on recent AAVSO observations. The color index of V503~Her varies between 1.15 and 1.45 mag as its brightness changes.

First, we confirmed that the median magnitudes of individual field stars were consistent between the ASAS $V$ and ASAS-SN $V$ datasets, as well as between the ASAS-SN $g$ and ZTF $g$ surveys, indicating that there were no additional shifts among the data obtained in the same bands. Then we obtained a shift of -0.59 mag from the $g$ to $V$ light curves for V503~Her, a result that was verified using overlapping observations in the ASAS-SN $V$ and $g$ light curves. For the DASCH data, calibrated using ATLAS-REFCAT2 \citep{2018ApJ...867..105T} to $g$ magnitudes, we found a shift of approximately -0.45 mag, but the lower precision of these measurements resulted in higher uncertainty. 
The NSVS and SuperWASP data were unfiltered and required separate analyses. We found a shift of +0.10 and +0.14 mag for the NSVS and SuperWASP data, respectively. These results were also confirmed by comparison with ASAS $V$ data obtained during the same period. Finally, we included observations from Sonnenberg plates \citep{1980VeSon...9..197M}, which have a sensitivity similar to that of the $B$ filter. We determined the relationship between $B-V$ and $B$ magnitude using recent multicolor AAVSO observations and used this relationship to convert observed $B$ measurements to $V$ magnitudes.

The resulting long-term light curve of V503~Her is shown in Fig. \ref{fig:lc_full}A. Although one needs to treat the absolute values with caution due to possible slight inaccuracies of these shifts and due to the fact that various filters have various effective wavelengths (at which the amplitude of variability can be different), this approach allows us to analyze the long-term variability of the object.

\subsubsection{Semi-regular pulsations}\label{sec:pulsations}

The most prominent feature of the V503~Her light curve is the semi-regular pulsations with variable amplitude on a variable timescale. These are especially well visible in the recent data, which were obtained with a~better cadence and precision than the historical observations (see Fig. \ref{fig:lc_full}B). Although the DASCH data do not have sufficient time sampling to estimate the amplitudes or the period of pulsations, the variance of the measurements around the median magnitude ($\sim12.78$\,mag in the original, unshifted dataset) suggests that pulsations were also present in the light curve during the time period covered by this dataset. The light curves constructed by \citet[][]{1980VeSon...9..197M} revealed quasi-periodic pulsations similar to those observed in the recent light curve, at least in certain time intervals.

Figs. \ref{fig:lc_full}C\&D show two segments of the recent light curve of V503~Her, demonstrating the variability in both the amplitude and timescale of the pulsations. Around JD\,2\,457\,100, the pulsations had a timescale of approximately 90 days and an amplitude of 0.65\,mag, while around JD\,2\,458\,300, the time difference between subsequent maxima increased to around 150 days, and the variability amplitude rose to 1.4\,mag. Between JD\,2\,458\,400 and JD\,2\,458\,600, the light curve showed a period of nearly constant brightness, with only minor variations of 0.15\,mag occurring on a timescale of 30-40 days. Previous observations based on Sonnenberg plates indicated the presence of periodicity within a range of 80-100 days \citep[][]{1980VeSon...9..197M}, while \citet{2013AcA....63..405G} derived a period of 130 days from ASAS data. The light curves of V503~Her in the VSX database \citep[][]{2006SASS...25...47W}, primarily based on NSVS data, show variability with a timescale of 112 days.

\begin{figure}
	\plotone{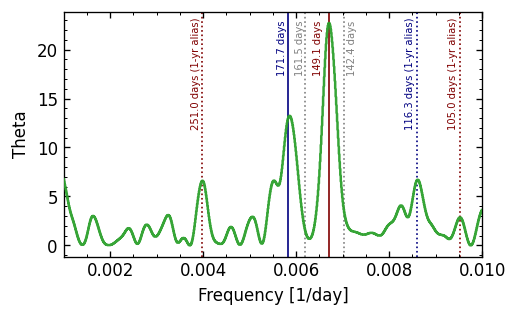}
        \plotone{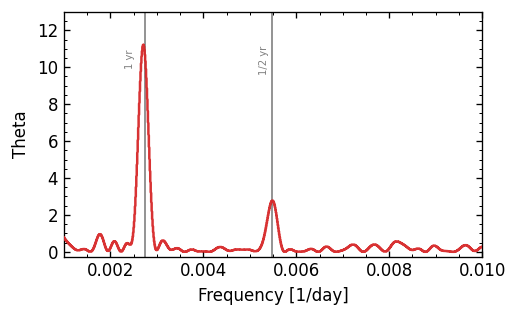}
    \caption{Periodogram (top panel) and spectral window (bottom panel) of the initial step of the period analysis of the V503~Her light curve based on the ASAS-SN survey data in the $V$ and $g$ filters. The most prominent periods (see Tab. \ref{periods}) and their 1-year aliases (assigned based on colors) are indicated by vertical lines with labels. Note that periods 142.4 and 161.5 days were only detected after removing the period responses found in the previous steps of the sequential analysis, and thus their peaks are overlaid by the nearby periods 149.1 and 171.7 days.}
    \label{fig:per_Vg}
\end{figure}

We conducted a comprehensive investigation of the pulsation behavior of V503~Her by performing a detailed period analysis of photometric observations of this object based on data from the ASAS-SN survey. Data sets in each filter ($V$, $g$) were analyzed separately as well as together. The preliminary period analysis was performed using the $Fo$ code based on Fourier harmonic analysis \citep{1994OAP.....7...49A}. Sequential analysis (with the implementation of the pre-whitening procedure between individual steps) of the observations in the $V$ filter yielded the most significant periods of 149.5 (and its 1-year aliases 105.4, 253.2), 158.1 (and its 1-year alias 110.9), 168.4 (and its 1-year alias 116.8), and 141.1\,days. Similarly, the most significant periods detected in the $g$~filter were 148.1 (and its 1-year alias 248.8) and 173.2 (and its 1-year alias 117.8) days. Analysis of the entire ASAS-SN light curve resulted in the detection of the most significant periods as 149.1 (and its 1-year aliases 105.0, 251.0), 171.7 (and its 1-year alias 116.3), 1\,100.5, 142.4, 161.5 days. The list of detected periods is given in the Tab. \ref{periods}. Periodogram and spectral window of the initial step of the period analysis of the V503 Her light curve based on the ASAS-SN survey data in the $V$ and $g$ filters is depicted in Fig. \ref{fig:per_Vg}. Slightly varying values (from $149.5 \pm 0.4$ to $148.1 \pm 0.5$ and from $168.4 \pm 0.9$ to $173.2 \pm 0.5$\,days) at different time periods may indicate changes in the period of the photometric variations of V503~Her.
Such a conclusion is also supported by the findings of the wavelet analysis (see below). 

\begin{table}
\caption{Detected periods determined by period analysis of photometric observations of V503~Her based on ASAS-SN survey data in the $V$ and $g$ filters. For each period $P$, the corresponding semi-amplitude $A_{1/2}$, the time of maximum $JD_{\rm max}$, and the significance $\theta/\langle\theta\rangle$ are given. Note that although the values given reflect quite well the relative significance of the detected periods to each other, they cannot be rigorously compared, since they were obtained in separate steps of the sequential period analysis with implementation of the pre-whitening procedure between individual steps.}
\label{periods}
\footnotesize
\begin{center}
\begin{tabular}{lcllc} 
\hline
$P$\,[days]	& $A_{1/2}$\,[mag] & $JD_{\rm max}$\,[day] & $\theta/\langle\theta\rangle$ & Filter \\
& & 24..& & \\
\hline
$149.5 \pm 0.4$	& $0.29 \pm 0.02$	& $57\,343.7 \pm 1.5$	& 81.8	& $V$	\\
$158.1 \pm 0.7$	& $0.16 \pm 0.02$	& $57\,290.9 \pm 2.7$	& 38.8	& $V$	\\
$168.4 \pm 0.9$	& $0.13 \pm 0.02$	& $57\,406.8 \pm 3.3$	& 31.8	& $V$	\\
$141.1 \pm 0.6$	& $0.11 \pm 0.01$	& $57\,387.8 \pm 2.6$	& 33.3	& $V$	\\
\hline
$148.1 \pm 0.5$	&  $0.22 \pm 0.02$	& $58\,984.8 \pm 1.7$	& 80.3	& $g$	\\
$173.2 \pm 0.5$	& $0.23 \pm 0.01$	& $58\,947.4 \pm 1.6$	& 109.6	& $g$	\\
\hline
$149.1 \pm 0.2$	& $0.22 \pm 0.01$	& $58\,388.8 \pm 1.4$	& 112.4	& $V$+$g$ \\
$171.7 \pm 0.3$	& $0.17 \pm 0.01$	& $58\,264.3 \pm 1.9$	& 87.9	& $V$+$g$	\\
$1\,100.5 \pm 14.2$ & $0.14 \pm 0.01$	& $58\,474.8 \pm 13.5$	& 68.1	& $V$+$g$	\\
$142.4 \pm 0.3$	& $0.11 \pm 0.01$	& $58\,375.2 \pm 2.3$	& 46.5	& $V$+$g$	\\
$161.5 \pm 0.4$	& $0.11 \pm 0.01$	& $58\,260.4 \pm 2.5$	& 50.8	& $V$+$g$	\\
\hline
\end{tabular}
\end{center}
\end{table}

\begin{figure}
	\plotone{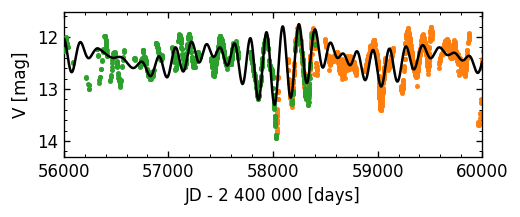}
    \caption{Model light curve of V503 Her (black) based on the analysis of ASAS-SN $V$+$g$ data (Tab. \ref{periods}). Colors of observed photometric datapoints in the light curve are the same as in Fig. \ref{fig:lc_full}.}
    \label{fig:beating}
\end{figure}

Interference of periods with similar values ($\sim$$149$, $\sim$$171$ days) could cause a light curve morphology with a visible beating pattern. The test superpositions of the responses with the obtained periods were able to partially fit the complex shape of the V503~Her light curve (see the example in Fig. \ref{fig:beating} with the periods obtained from the analysis of ASAS-SN $V$+$g$; Tab. \ref{periods}), but in neither case were they able to describe all its features, suggesting that the pulsation behavior is more complex.

\begin{figure}
	\plotone{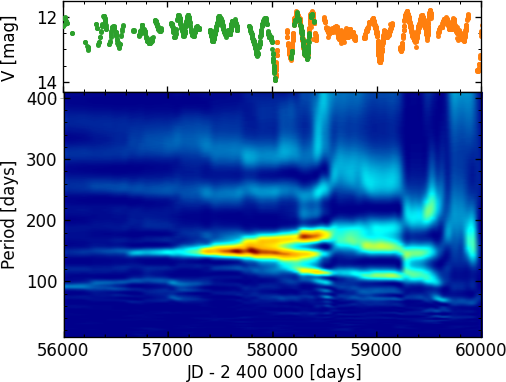}
    \caption{Results of Weighted Wavelet Z-Transform analysis of ASAS-SN data of V503~Her performed in software $Peranso$. Colors denote the wavelet power with dark blue having the lowest power and dark red having the highest. Colors of photometric datapoints in the corresponding light curve shown in the top panel are the same as in Fig. \ref{fig:lc_full}.}
    \label{fig:wwz}
\end{figure}

To conduct a more thorough period analysis of the light curve of V503~Her, we employed the Weighted Wavelet \mbox{Z-Transform} method \citep[WWZ; ][]{1996AJ....112.1709F} implemented in the $Peranso$ software \citep{2016AN....337..239P}. The resulting WWZ Period Window plot (Fig. \ref{fig:wwz}) confirms our preliminary findings that in the period under study (JD\,2\,456\,004 - 2\,459\,972) the photometric observations of V503~Her exhibit the presence of multiple periods (with dominant ones being around 90, 150, and 171\,days) with varying values at different epochs. These may not be present at the same time, nor may they be permanent, and the presence of other periods is not excluded. The superposition of the periods can lead to the observed beating appearance of the light curves (changes in the amplitudes of the photometric variations) with a period around 1\,000\,days. It should be noted that the value of this period is close to the time interval between deep minima around JD\,2\,458\,031, 2\,459\,032 and JD\,2\,459\,960.

The photometric behavior of V503~Her is typical for the SRb subtype of semiregular variable stars, characterized by semi-regular pulsations with variable amplitude and timescale. Typically, two (or less often three) radial modes of pulsations are detected in these objects. The dominant periods correspond to the pulsation in the fundamental and first overtone modes \citep[these lie on C and C' sequences in the period-luminosity diagram; see, e.g.,][and references therein]{2017ApJ...847..139T}. In most semi-regular pulsators, the ratios of the pulsation periods range from about 0.5 to 0.7 \citep{2013ApJ...779..167S}. A small group of SRb variables has dominant periods that lie between the C and D sequences \citep{2013ApJ...763..103S}, which also seems to be the case of V503~Her (for the inferred periods and Weisenheit index $W_{\rm JK} = K_{\rm S} - 0.686(J-K_{\rm S})$ = 12.45, scaled to the distance to the LMC of 49.59\,kpc, \citealt{2019Natur.567..200P}).

Unlike most SRb stars, which are typically giants of late spectral types (M, C, and S), the pulsating star in V503~Her is likely hotter, with a spectral type of K based on optical spectra (see below). This would place V503~Her in the subclass SRd. Based on the absolute $V$ magnitude calculated for a distance of 10.6 kpc, this component is more luminous than a giant of luminosity class III and lies close to class II, reaching even between classes II and Ib at maximum light \citep[][]{2013JKAS...46..103S}. This is consistent with the fact that semiregular variables are typically asymptotic giant branch stars (AGB) or red supergiant pulsating stars. The position of V503~Her in the period-luminosity diagram confirms its classification as an AGB star \citep[fig. 3 in][]{2017ApJ...847..139T}.

\subsubsection{Eclipsing nature}\label{sec:eclipses}

Based on the analysis of ASAS data, \citet{2013AcA....63..405G} suggested that V503~Her could be an eclipsing system. They proposed that two minima at JD\,2\,453\,145 and JD\,2\,454\,720 are eclipses, resulting in a possible orbital period of 1\,575 days. However, the new data collected in this study allowed us to detect several other minima in the light curve of V503~Her, during which the brightness decreased to the same level as during the second supposed eclipse (13.3 - 13.4\,mag in the \textit{V} filter; see Fig. \ref{fig:lc_full}B). These variations are caused by the pulsations (discussed above) as they occur on variable timescales, which cannot be expected for orbital variability. Therefore, even if it is a binary system, we consider the value of 1\,575 days for the orbital period of V503~Her to be incorrect.

There are three drops in brightness that are deeper than the regular pulsations, with $V$ magnitudes decreasing to $\sim$14.0 - 14.2\,mag around JD\,2\,453\,145 (the year 2004) and JD\,2\,458\,031 (the year 2017), and to $\sim$13.7\,mag around JD\,2\,459\,960 (the year 2023). We investigated the possibility that these drops are caused by eclipses. However, finding a single orbital period that would fit all three minima is impossible, and therefore, it is more likely that these minima are caused by the beating of the multi-periodic pulsations of V503~Her. Moreover, at least the minima in 2017 and 2023 seem to follow the timing of the pulsations rather well. The remote possibility that only the first two minima are caused by eclipses while the third by pulsations is discussed in the Appendix \ref{sec:eclipsing} for completeness.

\subsection{Stellar parameters of V503~Her}\label{sec:parameters}
In this section, we discuss the results of the analysis of optical spectra and SED of V503~Her, its stellar parameters, and evolutionary status.
\subsubsection{Optical spectra}

\begin{figure}
	\centering
    \plotone{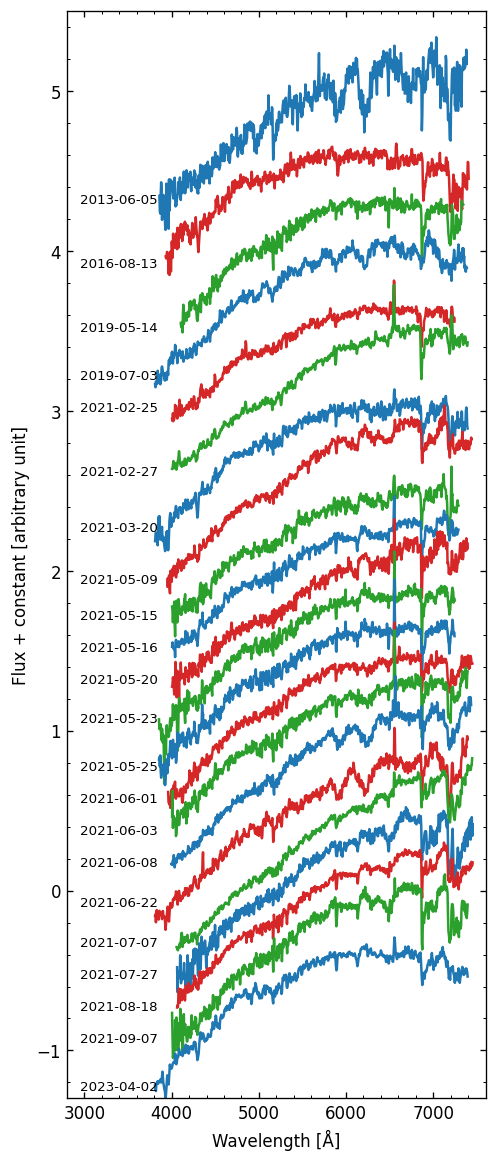}
    \caption{Optical spectra of V503~Her obtained by the ARAS Group.}
    \label{fig:spectra}
\end{figure}

Fig. \ref{fig:spectra} shows 22 low-resolution spectra of V503~Her, obtained between June 2013 and September 2021 (excluding one spectrum covering only the region redward of 6\,305\,\AA). The majority of our spectra exhibit a K-type star continuum, with mild TiO bands occasionally visible. The most prominent TiO bands are present in the spectra acquired in June 2013 and June 2021. These changes in spectral appearance are linked to photometric variations discussed in Sec. \ref{sec:long-term}. 

\begin{figure}
	\plotone{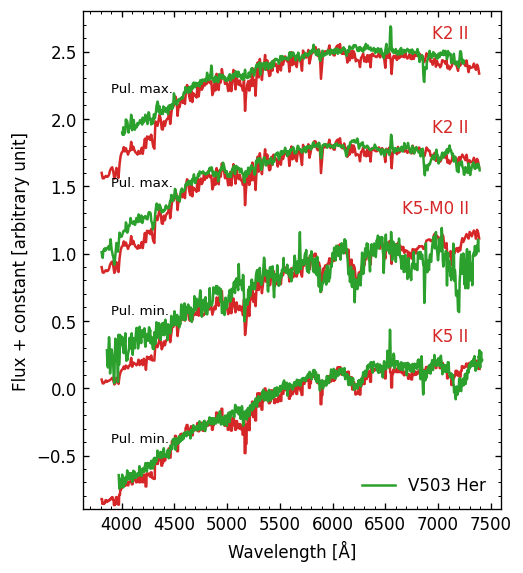}
    \caption{Low-resolution spectra of V503~Her (green color) obtained in various photometric phases (pulsation maxima in 2021 and 2023, pulsation minima in 2013 and 2021 acquired at JD\,2\,459\,270, 2\,460\,037, 2\,456\,448, and 2\,459\,373, respectively). The best-fitting empirical spectra from the MILES library \citep{2011A&A...532A..95F} are shown in red.}
    \label{fig:spectral_fit}
\end{figure}

The obtained spectra cover various phases of the pulsation of V503~Her (see the arrows in Fig. \ref{fig:lc_full}B), allowing a~more detailed analysis. To obtain the spectral classification, we compared the observed spectra with those from the MILES empirical library of stellar spectra of \citet{2011A&A...532A..95F}. The empirical spectra were down-sampled to the resolution of our spectra. At the photometric maximum, the optical spectrum of V503~Her resembles the K2 bright giant (Fig. \ref{fig:spectral_fit}). During the pulsation minimum, the TiO bands started to be prominent in the spectra, suggesting a decrease in the temperature of the pulsating star. Its spectrum resembles a~K5-M0 star. It is worth noting that the single star model does not seem to provide a satisfactory fit to all spectral features in some of the obtained spectra, e.g., see the exceeding flux around 5\,000\,\AA\,\,in that from 2013 and the excess in the blue part of some spectra (Fig. \ref{fig:spectral_fit}). These observations suggest that there may be an additional source of radiation present in V503~Her. However, given the low S/N values in the blue region of our spectra, one should interpret this finding with caution.

Our spectra are also consistent with the spectroscopic observation of V503~Her obtained by \citet{1978PASP...90..526B}. Although it is not clear from his paper when exactly the spectrum was obtained, according to the author's description, it showed well-developed TiO bands (similar to the M2 star but weaker) and also indicated the presence of a hotter star. This depiction fits well with our spectra obtained close to the pulsation minima of V503~Her, which show an excess in the blue part. Given that there are even deeper minima than those covered by our spectra observed in the light curve occasionally, it is possible that the spectrum published in Bond's article was acquired in such a photometric phase. On the other hand, the spectrum of \citet{2002A&A...383..188M} appears to be obtained close to the pulsation maximum when V503~Her shows a continuum of a K-type star.

\subsubsection{Multi-frequency SED analysis}\label{sec:SED}

\begin{figure}
  \plotone{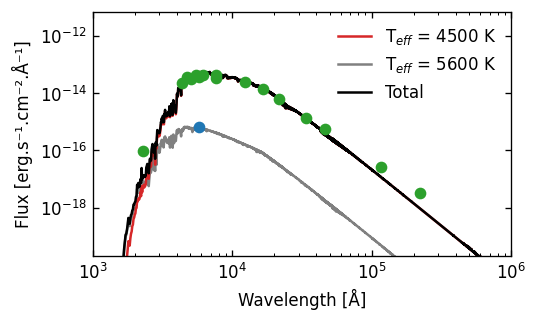}
    \caption{Multi-frequency SED of V503~Her (green points) constructed on the basis of data from the GALEX satellite, \textit{Gaia} EDR3, Pan-STARRS, APASS, 2MASS, and WISE. The best-fitting theoretical spectrum \citep{2014IAUS..299..271A} is shown in red. The spectrum shown in grey corresponds to the collective radiation of two background sources calibrated to its combined flux in the \textit{Gaia G} filter (blue point). See the text for more details.}
    \label{fig:sed_V503}
\end{figure}

During most of our observations, the optical spectra resemble a K bright giant star. The star appears cooler during the deep minima in the light curve, with mild TiO bands appearing in the spectra. To confirm the results obtained in the previous section, we constructed the multi-frequency SED of V503~Her using data from the GALEX satellite, \textit{Gaia} EDR3, Pan-STARRS, APASS, 2MASS, and WISE (Fig. \ref{fig:sed_V503}). The observed fluxes were de-reddened using the extinction value of $E(B-V) = 0.08$\,mag (Sec. \ref{sec:distance}). One should keep in mind that the data used were not obtained simultaneously, and therefore might not be acquired in the same pulsation phase, causing some scatter in the SED. On the other hand, in most cases, the fluxes taken from the particular catalogs are median values over several observing epochs, which reduces the impact on the results. Moreover, the use of data from several surveys observing at similar wavelengths also supports the consistency of the results.

To obtain the parameters of V503~Her, especially the effective temperature and luminosity, the SED was fitted with the BT-Settl grid of theoretical spectra \citep{2014IAUS..299..271A} using the VO SED analyzer (VOSA) on the Spanish Virtual Observatory theoretical services website \citep{2008A&A...492..277B}\footnote{http://svo2.cab.inta-csic.es/theory/vosa/}. The spectrum that best fitted the SED was the one with an effective temperature of 4\,500$\pm$50\,K, corresponding to the spectral type K2-K3 \citep[irrespective of the luminosity class; ][]{2009MNRAS.394.1925V}, which is well consistent with the results obtained from the comparison of optical spectra with empirical ones (K2 and K5-M0 during maximal and minimal brightness, respectively). We should note that the values of surface gravity $\log g$ = -0.5$\pm$0.25 and metallicity [Fe/H] = -1.5$\pm$0.25 obtained from the SED fit are significantly less well constrained. 

The best-fitting model was also used to infer the total observed flux of the source. Utilizing the distance estimate from Sec. \ref{sec:distance}, the luminosity of V503~Her, 1\,907$\pm$377\,L$_\odot$ can be estimated. From the Stefan-Boltzmann law, one can then infer the radius of 109$\pm$10\,R$_\odot$. The obtained value of luminosity is consistent with the luminosity class II suggested by the $V$ magnitudes (Sec. \ref{sec:pulsations}), as the tabulated values of luminosities for the K2\,II and M0\,II stars are 1\,400\,L$_\odot$ and 2\,340\,L$_\odot$, respectively \citep[][]{1987A&A...177..217D}.

\begin{figure}
    \plotone{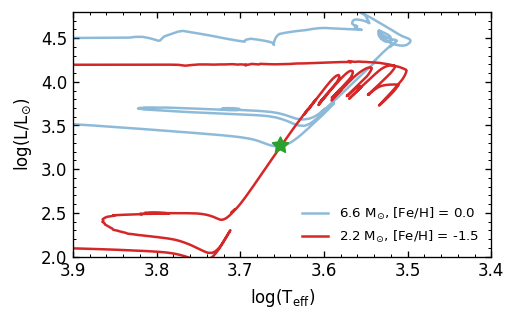}
    \caption{Position of V503~Her in the HR diagram for the obtained effective temperature 4\,500$\pm$50\,K and luminosity 1\,907$\pm$377\,L$_\odot$. The stellar evolutionary tracks shown in red and blue are from MIST \citep{2016ApJS..222....8D,2016ApJ...823..102C}.}
    \label{fig:hr}
\end{figure}

We have used the parameters obtained for V503~Her to investigate the evolutionary status of the star. The effective temperature and luminosity were compared to the MIST stellar evolutionary tracks \citep{2016ApJS..222....8D,2016ApJ...823..102C}, computed with the Modules for Experiments in Stellar Astrophysics (MESA) code \citep{2011ApJS..192....3P,2013ApJS..208....4P,2015ApJS..220...15P,2018ApJS..234...34P,2019ApJS..243...10P,2023ApJS..265...15J}. The resulting HR diagram is shown in Fig. \ref{fig:hr}. For the metallicity of -1.5 inferred from the SED, the best-fitting evolutionary track is that of a star with an initial mass of 2.2\,M$_\odot$. In such a case, the observed parameters are consistent with the AGB star, confirming our results based on the photometric variability (Sec. \ref{sec:pulsations}). Comparison with the evolutionary track for solar metallicity led to a significantly higher initial mass of the star ($\sim$6.6\,M$_\odot$) and suggested that the star is only about to start climbing the red giant branch. However, this seems to be contradicted by the observed variability. Therefore, the model with lower metallicity is preferred. Interestingly, all K-type giants in yellow symbiotic stars analyzed so far have sub-solar metallicities \citep[see, e.g., tab. 2 in ][and references therein]{2017ApJ...841...50P}. 

The SED also confirmed the presence of UV excess. According to the data from \textit{Gaia} DR3, there are two faint stars (Gaia DR3 4557410314849153280 and Gaia DR3 4557410314849154176) located in the vicinity of V503~Her that could contaminate the measured flux, especially in the GALEX \textit{NUV} filter given the angular resolution of the data. These stars can form a binary system since they are located at the same distance in front of V503~Her, have opposite proper motions, very similar $G$ magnitudes, and $G_{\rm BP} - G_{\rm RP}$ colors. Therefore, we investigated the possibility of whether the UV excess could be caused by these stars. We modeled both stars with a single theoretical spectrum with the effective temperature obtained from the $G_{\rm BP} - G_{\rm RP}$ color index using the relations of \citet{2021MNRAS.507.2684C}, calibrated to their combined $G$ magnitude. Our results show that while they apparently contribute to the UV flux (Fig. \ref{fig:sed_V503}), the excessive radiation in the GALEX \textit{NUV} filter cannot be explained solely by the flux of these two foreground stars. Another possible explanation is, therefore, that the UV radiation is intrinsic to V503 Her, as discussed in more detail in the next section. Unfortunately, the GALEX observations themselves are insufficient to constrain the parameters of the additional radiation source from the SED modeling with reasonable precision.

\subsection{Symbiotic classification}

Finally, having analyzed all the available observational data on V503~Her, we can discuss its proposed symbiotic classification. The object has been considered a~potential symbiotic binary since 1980 \citep{1980VeSon...9..197M}. Given the distance adopted, our results suggest the presence of a bright giant in V503~Her that meets the request for an evolved cool star in a potential symbiotic system. However, definitive confirmation of the nature of so-called shell-burning symbiotic stars requires optical spectra that satisfy several criteria. Specifically, in addition to late-type giant absorption features, they must exhibit strong emission lines of H I and He I, together with other emission lines with an ionization potential of at least 35\,eV and an equivalent width exceeding 1\,\AA\,\,\citep{1986syst.book.....K,2000A&AS..146..407B}. On the contrary, accreting-only symbiotic systems show weak or no emission lines in their optical spectra but are bright in UV and/or \mbox{X-rays} and exhibit short-term variability at these wavelengths \citep[][]{2019arXiv190901389M,2021MNRAS.505.6121M,2021PhDT........17L}.

The only emission lines detectable in our spectra of V503~Her are H\,I Balmer lines, particularly H$\alpha$. In a~minority of the spectra, H$\beta$ is also observed (Fig. \ref{fig:spectra}). For this reason, the classification of V503~Her as a shell-burning symbiotic system is ruled out. There remains the possibility that V503~Her is an accreting-only symbiotic system. This would be supported by the UV/near-UV observations of V503~Her, especially by its detection in the \textit{NUV} filter ($NUV$ = 20.44 $\pm$ 0.22 mag; $\lambda_{\rm eff}$ = 2305\,\AA) of the \textit{GALEX} satellite. A cool star like the one in V503~Her (see Sec. \ref{sec:parameters}) would have an intrinsic photospheric emission too faint to be detected by \textit{GALEX}. Therefore, UV detection suggests that there is probably another component in V503~Her or a source in the foreground/background. We also note that some of our optical spectra show an excess in the blue part, which would support the presence of an additional source of radiation with a higher temperature than that of K bright giant in V503~Her.

The investigated object was not detected in the \textit{FUV} filter ($\lambda_{\rm eff}$ = 1549\,\AA). However, given that the typical \textit{FUV-NUV} color of symbiotic stars is in the range of -1 to 2 \citep{2019RNAAS...3...28M,2023MNRAS.519.6044A} and even with only a~small extinction in the direction of V503~Her, the \textit{FUV} flux could be below the typical sensitivity limit of the \textit{GALEX} satellite \mbox{\citep[20 -- 21 mag; ][]{2014AdSpR..53..900B}}. About one year later, V503~Her was not detected in the \textit{UVM2} filter of \textit{Swift} \citep{2013A&A...559A...6L}, which has an effective wavelength similar to the \textit{NUV} filter of \textit{GALEX} ($\lambda_{\rm eff}$ = 2246\,\AA). Given that the brightness limit during the observations was about 1.5 mag fainter \mbox{(UVM2 $\sim$ 21.95 mag)}, it is likely that the UV source is variable. 

\begin{figure}
	\plotone{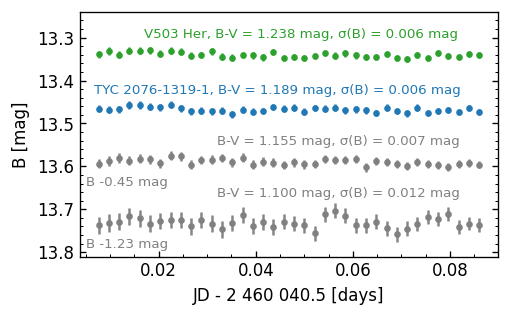}
    \caption{Short-term light curves of V503~Her (green), TYC~2076-1319-1 (blue), and two other fainter stars (shown in gray) in the vicinity obtained in the $B$ filter. The text in the figure gives the $B-V$ color of the particular star and the standard deviation of the data. Note that the light curves of fainter stars have been shifted by a constant shown in the figure for clarity.}
    \label{fig:flickering}
\end{figure}

If V503~Her is indeed an accreting-only symbiotic star, it could be somewhat similar to SU Lyn \citep{2016MNRAS.461L...1M,2021MNRAS.500L..12K,2021MNRAS.505.6121M,2022MNRAS.510.2707I} or recently discovered THA 15–31 \citep{2022A&A...661A.124M}. Although both objects host cooler red giants of M~spectral type, they show a strongly variable UV excess (see fig. 1 in \citealt[][]{2021MNRAS.505.6121M} for SU Lyn and fig. 1 in \citealt{2022RNAAS...6...54M} for observations of THA~15–31). SU~Lyn has recently been classified as a 'transient' symbiotic star by \citet{2022MNRAS.510.2707I}, showing characteristics of a stronger symbiotic interaction only at certain epochs. On the other hand, unlike SU Lyn or other accreting-only symbiotic stars, V503~Her was not detected as an X-ray source \citep{2013A&A...559A...6L,2020ApJS..247...54E}. However, the non-detection might be just due to either the limited sensitivity of the current X-ray missions, in particular to the targets at such a distance \citep{2021MNRAS.505.6121M}, and/or V503~Her might have been in some low state during the pointings. The fact that the target was not detected in UV during the \textit{Swift} observation would support such a hypothesis.

Motivated by a possible accreting-only symbiotic nature of V503~Her, we executed a 2-hour-long observing run on the Mayer 65-cm telescope in Ondřejov, Czech Republic, to search for possible signatures of flickering in its light curve. Given the issues related to the sensitivity of modern CCD cameras in the $U$ filter and possible complications caused by frequent contamination of the data obtained in most $U$ filters by photons from the red region \citep[see the discussion in][]{2021MNRAS.505.6121M}, we obtained the data for the flickering study in the Johnson $B$ filter. The obtained light curve of V503~Her is shown in Fig. \ref{fig:flickering}, together with light curves of three nearby stars with similar color and brightness as close as possible. The selection of stars with close color and brightness ensures minimal influence of unequal atmospheric transmission and statistical noise. The data for V503~Her do not indicate the presence of short-term flickering-like variability with a larger amplitude than the noise that affects nearby stars. However, one should still keep in mind that even in the case of well-known accreting-only symbiotic stars (e.g., SU~Lyn, MWC~560, or RT Cru), flickering is not detectable at all epochs \citep[e.g., ][]{2020MNRAS.492.3107L,2021MNRAS.505.6121M,2022BlgAJ..37...62G,2023BlgAJ..38...83Z,2023A&A...670A..32P}, so the non-detection of flickering does not reject the accretion-only symbiotic nature of V503~Her.

Given that the amplitude of the flickering typically increases toward blue, if it is present in V503~Her, it might be more easily detectable in UV. Observations in this wavelength range, together with repeated X-ray observations at the time when V503~Her would be in a high state (e.g., triggered by the detection of the strong UV excess in optical spectra), could provide additional evidence towards the accreting-only symbiotic classification of V503~Her.

To complete the discussion, we should add that V503 Her does not meet the IR criteria for the classification of S-type symbiotic stars nor would be classified as a symbiotic star in most of the classification trees as recently proposed by \citet{2019MNRAS.483.5077A,2021MNRAS.502.2513A}. However, these criteria cannot be used alone to confirm or reject the symbiotic classification of an object. Still, for now, the detection of UV radiation from the target seems to be the only evidence pointing towards the symbiotic nature of V503 Her.

In addition to binarity (accreting-only classification), there is possibly also another explanation for the UV emission of V503 Her. Recent efforts showed that a significant fraction of galactic AGB stars show UV emission \citep[e.g.,][]{2017ApJ...841...33M,2022Galax..10...62S}. Although in several cases the emission is claimed to arise from binarity, other models attribute the UV radiation to the chromospheres of AGB stars. The binary model is preferred, in particular, in the cases where the FUV emission is detected or when the observed-to-predicted ratio for the NUV emission is very high \citep[especially for the main-sequence companions; see][]{2016MNRAS.461.3036O,2019MNRAS.482.4697O,2022Galax..10...62S}, which is not the case for V503 Her. The less pronounced UV excess can be explained by active chromosphere models \citep[see][]{2022Galax..10...62S}. \citet{2017ApJ...841...33M} found that NUV emission of some AGB stars is correlated with optical and NIR light curves, suggesting that it is connected to the star itself. Moreover, \citet{2020MNRAS.491..680G} also found a correlation between pulsations of large amplitude and UV characteristics.

As discussed above, the observations of V503~Her obtained by \textit{GALEX} and \textit{Swift} suggest that UV emission is variable. The two observations were carried out about one year apart, the \textit{GALEX} observing the target at the maximum optical brightness, while the \textit{Swift} at the time when the brightness was decreasing after the pulsation maximum (but not at the minimum; the difference in $V$ was only about 0.6\,mag). Our spectroscopic observations do not allow us to search for signatures of chromospheric activity, e.g., the emission component in the Ca II H and K doublet, as the resolution of our spectra is low, as well as the S/N at that part of the spectrum is very poor in our data. In general, to fully distinguish between binarity and chromospheric activity, UV spectroscopy is needed \citep[see, e.g., the case of Y Gem][]{2018ApJ...860..105S}, which is unfortunately not available for V503 Her. 
\vspace{-3mm}
\section{Conclusions}
\label{sec:conclusions}
V503~Her has been classified as a symbiotic candidate since 1980. In some databases (e.g., SIMBAD), it is even considered a~confirmed symbiotic binary, although no detailed study of this object has been presented in the literature. Based on the detection of two minima in the ASAS light curve, it was also classified as an eclipsing binary system. In this study, we have investigated extensive long-term photometric observations obtained by various surveys, SED of the object, and its newly obtained optical spectra. Our main findings can be summarized as follows: 

\renewcommand{\theenumi}{\alph{enumi}}
\begin{enumerate}
    \item the light curves of V503~Her are dominated by the semi-regular multi-period pulsations whose interference causes very complex patterns,
    \item most probably, the object is not an eclipsing binary system as the eclipse-like minima can be satisfactorily explained by the presence of pulsations,
    \item the optical spectra show the K-type continuum with mild TiO bands appearing in certain pulsation phases,
    \item the pulsating star is a K-type bright giant on the asymptotic giant branch located at the distance of $\sim$\,10.6\,kpc with an effective temperature of 4\,500\,K, luminosity of 1\,900\,L$_\odot$, and sub-sollar metallicity,
    \item V503~Her cannot be classified as a shell-burning symbiotic system, while some pieces of evidence (in particular the variable UV emission) support its classification as an accreting-only symbiotic star,
    \item the active chromosphere of AGB star might be also responsible for the observed UV emission of V503 Her.
\end{enumerate}

Further multi-frequency observations of V503~Her are encouraged in order to definitively establish its symbiotic nature. The target is especially interesting as it might belong to a currently very small group of yellow accreting-only symbiotic systems. Furthermore, in the scope of the evolution of the giant in V503~Her, the mass loss would probably increase significantly (as predicted by evolutionary models), which could lead to a stronger interaction with the possible secondary component and the persistent appearance of symbiotic behavior in the future.


\begin{acknowledgments}
\small We are thankful to an anonymous referee for the comments and suggestions improving the manuscript. We are grateful to F. Teyssier for coordinating the ARAS Eruptive Stars Section. This research was supported by the \textit{Charles University}, project GA UK No. 890120, and \textit{Slovak Research and Development Agency} under contract No. APVV-20-0148. 

This publication makes use of VOSA, developed under the Spanish Virtual Observatory (https://svo.cab.inta-csic.es) project funded by MCIN/AEI/10.13039/501100011 033/ through the grant PID2020-112949GB-I00. VOSA has been partially updated using funding from the European Union's Horizon 2020 Research and Innovation Programme under Grant Agreement no. 776403 (EXOPLANETS-A). The DASCH project at Harvard is grateful for partial support from NSF grants AST-0407380, AST-0909073, and \mbox{AST-1313370}.
\end{acknowledgments}

%

\vspace{-0mm}


\software{\small astropy \citep{2013A&A...558A..33A,2018AJ....156..123A,2022ApJ...935..167A}, matplotlib \citep{Hunter:2007},
NumPy \citep{harris2020array},
PyAstronomy \citep{pya},
SpectRes \citep{2017arXiv170505165C},
          Peranso \citep{2016AN....337..239P}, 
          Fo \citep{1994OAP.....7...49A},
          VOSA \citep{2008A&A...492..277B},
          MIST \citep{2016ApJS..222....8D,2016ApJ...823..102C},
          MESA \citep{2011ApJS..192....3P,2013ApJS..208....4P,2015ApJS..220...15P,2018ApJS..234...34P,2019ApJS..243...10P,2023ApJS..265...15J}
          }



\begin{table*}[h!]
\appendix
\noindent\mbox{}\vrule height 24pt width0pt\hfill{\apjsecfont
APPENDIX}\hfill\mbox{}\par 
\vskip5pt
\section{Log of observations}\label{app:log}

\caption{Log of spectroscopic observations. Observer codes: BUI = C.~Buil, CBO = C.~Boussin, DBO = D.~Boyd, FAS = F.~Sims, JCO = J.~P.~Coffins, JMO = J.~Montier, JRF = J.~R.~Foster, PAD = P.~A.~Dubovský, TME = T.~Medulka.}             
\label{table:log_obs}      
\centering
\begin{tabular}{llclllcl}
\hline\hline
JD& Res. & $\lambda_{\rm min}$-$\lambda_{\rm max}$ & Obs. & JD& Res. & $\lambda_{\rm min}$-$\lambda_{\rm max}$ & Obs.\\
2\,4.. &  & [\AA] & & 2\,4.. &  & [\AA] & \\\hline
56\,448.884 & 603 & 3856-7391 & BUI & 59\,357.272 & 1031 & 3850-7251 & FAS \\
57\,613.979 & 654 & 3911-7396 & JMO & 59\,359.242 & 1056 & 3851-7250 & FAS \\
58\,617.942 & 1080 & 4101-7350 & DBO & 59\,366.915 & 881 & 3900-7550 & TME, PAD \\
58\,667.361 & 653 & 3720-7391 & JRF & 59\,368.899 & 889 & 4000-7550 & TME, PAD \\
58\,670.350 & 905 & 6305-9446 & JRF & 59\,373.927 & 978 & 3950-7550 & PAD \\
59\,270.531 & 792 & 4001-7251 & FAS & 59\,387.324 & 544 & 3651-7400 & JRF \\
59\,272.194 & 545 & 4000-7500 & JCO & 59\,402.004 & 1121 & 4000-7500 & PAD \\
59\,293.170 & 523 & 3750-7566 & CBO & 59\,422.980 & 1104 & 4000-7500 & PAD \\
59\,343.959 & 921 & 3950-7550 & PAD & 59\,444.867 & 1037 & 4000-7500 & PAD \\
59\,349.265 & 994 & 4000-7296 & FAS & 59\,464.885 & 1068 & 4000-7550 & PAD \\
59\,350.299 & 1066 & 4000-7295 & FAS & 60\,037.032 & 544 & 3625-7400 & JRF \\
59\,354.913 & 977 & 4000-7550 & PAD &  &  &  &    \\\hline
\end{tabular}
\end{table*}
\begin{table*}[]
\caption{List of the archival photometric observations used in this paper.}             
\label{table:log_obs_photometric}      
\centering
\begin{tabular}{llcl}
\hline\hline
Survey & Filter(s) & Range of JD - 2 400 000 {[}days{]} & Range of dates  \\\hline
DASCH & $g$ & 23\,166 - 47\,466 & April 22, 1922 - October 31, 1988 \\
\citet{1980VeSon...9..197M} & $B$ & 27\,542 - 41\,183 & April 14, 1934 - August 19, 1971 \\
NSVS & unfiltered & 51\,273 - 51\,623 & April 4, 1999 - March 19, 2000 \\
ASAS & $V$ & 52\,032 - 54\,836 & May 2, 2001 - January 4, 2009 \\
AAVSO & $B$, $V$ & 52\,067 - 59\,972 & June 7, 2001 - January 28, 2023 \\
SuperWASP & unfiltered & 53\,128 - 54\,325 & May 3, 2004 - August 13, 2007 \\
ASAS-SN & $V$, $g$ & 56\,004 - 59\,978 & March 17, 2012 - February 3, 2023 \\
ZTF & $g$ & 58\,203 - 59\,972 & March 25, 2018 - November 7, 2022   \\\hline
\end{tabular}
\end{table*}
\begin{table*}[]
\caption{Photometric data used for the analysis of SED. For the APASS survey, the observing date is not available in the catalog.}             
\label{table:log_obs_sed}      
\centering
\begin{tabular}{llcl}
\hline\hline
Survey & Filter(s) & Median JD - 2 400 000 {[}days{]} & Median obs. date  \\\hline
\textit{GALEX} & \textit{NUV} & 54\,980 & May 26, 2009 \\
\textit{Gaia} DR3 & \textit{$G_{\rm BP}$, \textit{G}, \textit{$G_{\rm RP}$}} & 57\,387 & December 31, 2015 \\
Pan-STARRS & \textit{g, r, i, z, y} & 55\,544 & December 13, 2010 \\
APASS & \textit{B, V, g', r', i'} & - & - \\
2MASS & \textit{J, H, K} & 51\,605 & March 01, 2000 \\
\textit{WISE} & \textit{W1, W2, W3, W4} & 55\,316 & April 29, 2010   \\\hline
\end{tabular}
\end{table*}

\clearpage
\appendix
\setcounter{section}{1}
\section{Can V503~Her still be a long-period eclipsing binary?}\label{sec:eclipsing}
As discussed in Sec. \ref{sec:eclipses}, the minima in the light curve of V503~Her seem to be caused by the complex multi-period pulsation pattern. However, we review here the remote possibility that two deeper features during which the $V$~magnitude decreased to \mbox{$\sim14.0 - 14.2$\,mag,} detected around JD\,2\,453\,145 and JD\,2\,458\,031 are caused by eclipses. 

Considering that both minima are primary eclipses and that there has been no other such eclipse between these events, we can provide the linear ephemeris:

\begin{equation}
\label{eq:eph}
JD{\rm_{min}} = 2\,458\,031 + 4\,883 \times E.  
\end{equation}

If the orbital period was half of the obtained value, another eclipse should have occurred around JD\,2\,455\,589. While this part of the V503~Her light curve is not well covered by data, there is its observation in the AAVSO database obtained at JD\,2\,455\,598 (nine days after supposed light minimum) during which the star had magnitude \mbox{$V = 12.04$\,mag}. That is much more than the expected value of \mbox{$V \sim 13.35 - 13.85$\,mag} (based on the two other minima). The discrepancy of $1.31 - 1.81$\,mag suggests that no eclipse occurred in this time period, and the only possible period is indeed that of 4\,883\,days. 
A~period with a duration of one third of the proposed value (or any shorter) is excluded since additional eclipses would have been detected in the recent \mbox{ASAS-SN} and AAVSO light curves of V503~Her. The possibility that one observed minimum is primary and the second one secondary is ruled out by the ratio of expected fluxes of possible primary and secondary components. With similar depths, they need to be similar, and consequently, the secondary would be detected in the optical spectra.

The orbital period of 4\,883\,days (13.4\,years) could make V503~Her an eclipsing binary with the sixth longest orbital period known. In total, there are only seven other eclipsing binaries (with known periods and observed eclipses) that have orbital periods longer than ten years in GCVS \citep[][]{2017ARep...61...80S}: AS~LMi \citep[25\,245\,days; ][]{2016A&A...588A..90L}, $\epsilon$~Aur \citep[9\,890\,days; ][]{2010IBVS.5937....1C}, VV~Cep \citep[7\,430\,days; ][]{1937HarCi.421....1G}, V381~Sco \citep[6\,545\,days; ][]{1940AnHar..90..231S}, $\gamma$~Per \citep[5\,330\,days;][]{1999A&A...348..127P}, V383~Sco \citep[4\,876\,days; ][]{2013A&A...550A..93G}, and V695~Cyg \citep[3\,784\,days; ][]{2008Obs...128..362G}. One or both components of all of these binaries is a giant or supergiant star. For completeness, we should add that a bright binary $\alpha$~Com, consisting of two dwarf stars with an orbital period of 9\,442\,days is predicted to show the eclipses \citep[][]{2010AJ....140.1623M}. However, they were not observed up to now. The eclipse predicted at the end of 2014 was missed due to miscalculations of the time of eclipse \citep[][]{2010AJ....140.1623M, 2015AJ....150..140M}, and therefore this target will be included in the list above only after successful observation of its next eclipse predicted in 2040.

\begin{figure}
	\plotone{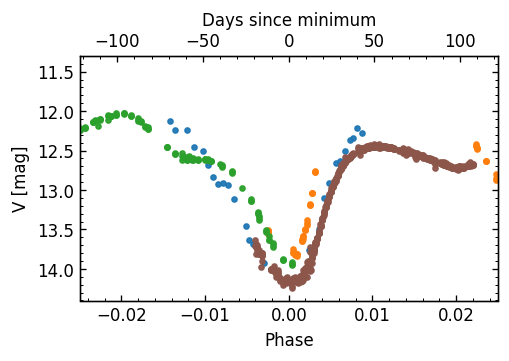}
    \caption{Two parts of the V503~Her light curve phased according to Eq. \ref{eq:eph}. The blue and brown points depict the minimum observed in 2004 by ASAS ($V$ filter) and SuperWASP (unfiltered, shifted by +0.14 mag to the level of the ASAS $V$ observations), respectively. The green and orange points depict the minimum observed in 2017 by ASAS-SN in $V$, and $g$ (shifted by -0.59 mag) filters, respectively.}
    \label{fig:eclipses}
\end{figure}

Two parts of the V503~Her light curve phased according to the ephemeris in Eq. \ref{eq:eph} are shown in Fig. \ref{fig:eclipses}. However, while it would be tempting to classify V503~Her as an eclipsing symbiotic system with the sixth longest known orbital period, several facts play against this object's eclipsing nature. The minima profiles are rather different from those expected for real eclipses. Especially, since the minimum brightness in both events is similar, one should expect that the radii of the eclipsing components are also similar during these two epochs. However, this does not seem to be the case, given the significant difference in the time intervals between the ingress and egress of the two events, leading to a rather different sum of the radii. 

The prominent variability in the V503~Her light curve makes it impossible to estimate the exact contact times of the supposed eclipses and use them to precisely determine the parameters of the components of the eventual binary system. Nevertheless, approximate estimates can still be made. Adopting the generalized Kepler's third law and the conservative estimate of the total mass of a~binary system between 2 and 5\,M$_\odot$ results in a~semimajor axis of about 7.0 - 9.5\,au. For the duration of the eclipse of $\sim 110$\,days (for the 2004 event) and the assumption of $i = 90^{\circ}$, we can obtain the sum of the radii of the supposed binary components $R_1 + R_2$ $\sim 96 - 166\,\rm R_\odot$. The duration of the 2017 event appears to be about twice as short, leading to the sum of the radii $R_1 + R_2$ $\sim 48 - 83\,\rm R_\odot$. 

As mentioned above, one would expect that the values would be similar during both minima, which is clearly not the case here. Moreover, these results do not seem to be consistent with the parameters (especially the radius) of the K bright giant that we inferred in Sec. \ref{sec:parameters}. Although the radius of the pulsating star would surely be slightly variable, its changes would not be sufficient to accommodate another star in the system if the sums of radii above are correct. We should note that the assumption of the inclination $i = 90^{\circ}$ makes these values only lower limits. However, although it is not possible to identify the total phase of the eclipses in the light curve, their presence in such a long-period binary requires the inclination to be very close to 90 degrees. 

To conclude, combining the analysis discussed in this appendix with the results obtained in the article's main body (in particular, the detection of other deep minima in the light curve and their timing), the eclipsing nature of V503~Her seems to be rather improbable.


\bibliography{sample631}{}
\bibliographystyle{aasjournal}



\end{document}